# Photoluminescence of hexagonal boron nitride: effect of surface oxidation under UV-laser irradiation

**Luc Museur***

*Laboratoire de Physique des Lasers – LPL, CNRS, Institut Galilée, Université Paris 13, 93430 Villetaneuse, France*

**Demetrios Anglos**

*Institute of Electronic Structure and Laser (IESL) FORTH  711 10 Heraklion Crete Greece*

**Jean-Pierre Petitet, Jean Pierre Michel and Andrei V. Kanaev**

*Laboratoire d'Ingénierie des Matériaux et des Hautes Pressions – LIMHP, CNRS, Institut Galilée, Université Paris 13, 93430 Villetaneuse, France*

* *Corresponding author: museur@galilee.univ-paris13.fr*

## Abstract:

We report on the UV laser induced fluorescence of hexagonal boron nitride (h-BN) following nanosecond laser irradiation of the surface under vacuum and in different environments of nitrogen gas and ambient air. The observed fluorescence bands are tentatively ascribed to impurity and mono ($V_N$), or multiple (m-$V_N$ with m = 2 or 3) nitrogen vacancies. A structured fluorescence band between 300 nm and 350 nm is assigned to impurity-band transition and its complex lineshape is attributed to phonon replicas. An additional band at 340 nm, assigned to $V_N$ vacancies on surface, is observed under vacuum and quenched by adsorbed molecular oxygen. UV-irradiation of h-BN under vacuum results in a broad asymmetric fluorescence at ~400 nm assigned to m-$V_N$ vacancies; further irradiation breaks more B-N bonds enriching the surface with elemental boron. However, no boron deposit appears under irradiation of samples in ambient atmosphere. This effect is explained by oxygen healing of radiation-induced surface defects. Formation of the oxide layer prevents B-N dissociation and preserves the bulk sample stoichiometry.

# I. Introduction

During the past years, electronic and crystalline structures of boron nitrides (BN) have been studied in detail, both theoretically and experimentally [1, 2]. This is mainly due to important properties of BN such as high electrical resistance and thermal conductivity, extreme mechanical hardness, elevated melting point, large band gap energy that make this material interesting for many large-scale applications in electronics, optoelectronics, protective coating, etc.

Graphite-like hexagonal BN (h-BN) is the most commonly used polymorph of boron nitride existing also as cubic (diamond-like), rhombohedral, and wurzitic. The electronic structure of h-BN has been studied by luminescence [3-8], optical reflectance and absorption [9-12], x-ray emission [13-15], inelastic x-ray scattering [16, 17], x-ray absorption [15, 18, 19], electron energy loss [20-22] spectroscopy. Despite extensive studies, up to now there is disagreement on its basic electronic properties. For example, both direct and indirect band gap natures and widely dispersed band gap energy values ranging from 3.6 eV to 7.1 eV have been reported in the literature [8]. The h-BN band structure theoretical calculations in the LDA approximation result in the lowest indirect gap around 4 eV [23-27], whereas recent GW calculations increase this value to 5.95 eV [27]. The effects of stacking on the electronic properties of h-BN has been studied [26] predicting the indirect band gap close to 4.0 eV for the most common form of h-BN.

The interest to luminescence properties of h-BN has been recently renewed with observation of laser effect at 215 nm in monocrystalline sample under e-beam excitation [4]. Although the photoluminescence (PL) of many III-V group semiconductors such InP, GaAs GaN etc. have been extensively studied in the past, only few data exist in the literature about the BN photoluminescence. With excitation at $\lambda$ = 262 nm Larach *et al.* [6] have observed a fine-structure luminescence in the range 300-500 nm and considered it to be inherent to the BN molecular layer. Katzir *et al*. [28] have observed a blue PL continuum in the range 390-500 nm ($\lambda_{exc}$ = 320 nm) attributed to deep levels of carbon impurities. Afterwards the effects of carbon doping on the blue luminescence has been studied by several groups [29-31]. Yao *et al*. [32, 33] have investigated the visible PL dependence on a degree of three-dimensional sample ordering.

So far the luminescent mechanism in h-BN is not clear and needs further investigations. A large diversity in shapes and positions of PL spectra published in the literature are usually interpreted on the basis of impurities presence and sample preparation conditions. Some important parameters as environment, excitation intensity, and light-induced changes in samples have not been yet considered. In the present article we report on room temperature photoluminescence of h-BN following 248-nm laser irradiation below ablation threshold. We show influence of irradiation dose, fluence and of environmental atmosphere on PL spectra, which allows distinguishing between surface and bulk fluorescent centers. We underline the importance of the BN surface oxidation induced by UV irradiation in the presence of oxygen gas.

These results are of interest in the frame of the current research on BN nanotubes (NT) which are more oxidation resistant and are expected to have better electronic properties than carbon NT [34]. Recently, the attempts[35, 36] to use PL measurements to recognize *in situ* BN nanotubes mixed with microcrystalline h-BN have been limited by the lack of precise knowledge about h-BN luminescence. Moreover the use of BN nanotubes as UV light source could be complicated by the possibility of surface oxidation.

## II. Experiment

Experiments were carried out at the I.E.L.S - F.O.R.T.H using a nanosecond ($\tau_L$ = 15 ns) KrF excimer laser ($\lambda$ = 248 nm). The samples were placed in a home made high pressure cell, with sapphire windows, which allows varying the pressure in the range from $10^{-2}$ mbar to 1 kbar.[37] The laser beam was focused on the sample surface at room temperature. The energy on the sample was change by using two reflective attenuators and was carefully measured using an energy meter and taking into account the beam geometry. Typical laser fluence in fluorescence experiments were between 1 and 100 mJ/cm². The laser induced fluorescence was collected by a quartz lens and focused on a multimode quartz optical fiber. A compact spectrograph equipped with a concave holographic grating, combined with an ICCD detector (DH520-18F, Andor Technology), was used to record the time-integrated emission spectrum from 200 to 800 nm with a resolution of about 2 nm.

In the experiments the UV laser has been used in two different modes of fluence. In the first mode, high fluence pulses (100mJ/cm$^2$) and long irradiation periods are used to possibly induce transformations on the sample. In this way dose irradiation of several 100J/cm$^2$ has been achieved. In the second mode, low fluence pulses (typically 2–3 mJ/cm$^2$) are used to probe the sample. Due to the accumulation of 500 luminescent pulses on the CCD detector, this corresponds to an irradiation dose of 1 J/cm$^2$. Except one, all the luminescent spectra presented in this article have been obtained with low fluence probe pluses.

The samples were square pallets (8x8x1 mm$^3$) compacted from hBN powder (Alfa, 99,5%) under the pressure of 0.6 GPa. To avoid organic impurities and traces of water, the pallets were heated at 800 K under vacuum for a period of 12 hours. The grit size of the hBN powder used for the experiment has been estimated by granulometry and transmission electron microscopy (JEM 100C JEOL). It ranged from 0.3 to 10 μm with an average particle size of 3.1 μm corresponding to the maximum in the mass distribution curve.

## III. Results and discussion

The h-BN is a highly fluorescent material under optical excitation above 4 eV [8, 38]. Its fluorescence spectra show different bands: (i) an asymmetric-shape continuum expanding into the visible with a maximum at 370 nm (UV1), (ii) a structured band with four maxima between 300 nm and 350 nm (UV2), and (iii) a broad visible continuum below ~400 nm (V1). The bands UV1 and UV2 have been earlier assigned correspondingly to the surface and bulk states emission. Their origins however are still debated. In our early studies [38] we have analyzed the effect of nanosecond laser irradiation (248 nm) on polycrystalline h-BN under vacuum. In short it is resumed in surface enrichment by elemental boron due to B-N bonds breaking [39]:

$$BN(s) + h\nu(5.0\,eV) \rightarrow B(s) + 1/2\,N_2(g) \qquad (1)$$

where (s) and (g) represent respectively solid and gas-phase species. This photochemical process requires an activation energy of 2.57 eV that is smaller than that of the UV laser photons. The surface modification has been spectroscopically observed by a decrease of the intensity of characteristic UV emission bands [38].

Below we will discuss the nature of h-BN fluorescence bands appearing following nanosecond laser surface irradiation in different environments. The observed spectral features will be tentatively ascribed to mono and multiple (di and tri) nitrogen vacancies: $V_N$, $2\text{-}V_N$, and $3\text{-}V_N$.

*Luminescence under vacuum and in ambient air*

Luminescence spectra of h-BN excited under ambient air and under vacuum conditions are shown on Fig.1. The shapes of the two spectra are different. In particular a new band labeled "V" appears around 340 nm in vacuum. The modification of the spectra is completely reversible and the band V disappears after the sample is exposed to ambient air. Moreover when the sample is irradiated in pure nitrogen gas (p=50 bars), the luminescence spectrum is identical to that obtained under vacuum. This indicates its quenching by oxygen. The band V can be visualized by taking the difference spectrum between that under vacuum and air. It is centered at ~340 nm with $\Delta\lambda_{hwfm}$ = 40 nm. Optical properties of h-BN are known to be strongly affected by nitrogen vacancies, which serve as activation, recombination, absorption and photosensitivity centers. Because of the oxygen healing effect and the energetic position blue shifted with respect to the transition ( $A \rightarrow S_c$ ) of bulk nitrogen vacancy [11], we assign the fluorescence band V to the surface nitrogen vacancies $V_N$ free of adsorbed oxygen molecules.

The oxygen present in the cell can quench the fluorescence in the following process:

$$V_N^* \rightarrow V_N + h\nu$$
$$V_N + O_2 \rightarrow \text{quenching} \qquad (2)$$

where $V_N^*$ refer to the fluorescent centre. Based on these results it is instructive to estimate the binding energy of the adsorbed $O_2$, which can clear up the site nature. The process (3) maintains the equilibrium between the free and adsorbed oxygen molecules.

$$O_2 + V_N \xleftrightarrow{R^-/R^+} V_N - O_2 \qquad (3)$$

The adsorption rate $R^+$ of $O_2$ molecules on the sample surface can be expressed as

$$R^+ = 1/4 \cdot \sigma V_T [O_2] \cdot (n_s - n) \qquad (4)$$

where $\sigma = 9.95 \cdot 10^{-16}$ cm² is gas-kinetic cross-section, $V_T = 3.9 \cdot 10^4$ cm/s its thermal velocity T=298 K, and [$O_2$] concentration of oxygen molecules, $n_S$ is the total concentrations of fluorescent sites, and n is the concentration of occupied by absorbed oxygen fluorescent sites. The desorption rate $R^-$ can be expressed as

$$R^- = \nu_D n \cdot \exp(-E^*/kT) \qquad (5)$$

where Debye frequency $\nu_D = k\theta_D/h = 8.33 \cdot 10^{12}$ Hz ($\theta_D$ =400 K is Debye temperature of h-BN), T is the ambient temperature, and $E^*$ is the binding energy. Assuming the equilibrium (3)-(5) and that fluorescence intensity $I_{fluo} \propto (n_s - n)$ one can obtain:

$$I_{fluo} \propto \frac{\frac{4\nu_D}{\sigma V_T} \exp(-E^*/kT)}{([O_2] + \frac{4\nu_D}{\sigma V_T} \exp(-E^*/kT))} \qquad (6)$$

The fit of the experimental data presented in inset of Fig. 1 by solid line results in $\frac{4\nu_D}{\sigma V_T} \exp(-E^*/kT) = 2.2 \pm 0.6 \cdot 10^{14}$ cm⁻³. From this value we can obtain the binding energy of oxygen molecules on the relevant surface site: $E^* = 0.61$ eV. This value is not large enough to account for chemisorption. Apparently, atmospheric oxygen is physisorbed at $V_N$ surface vacancies in the molecular form.

### *Effect of irradiation dose*

As we have already remarked, 5.0-eV photons produce an enrichment of h-BN by elemental boron (1). This effect, which has been observed under vacuum at laser doses ≥10 J/cm$^2$ [38] , is accompanied by a darkening of the surface and a complete disappearance of the characteristic fluorescence from the irradiated area at long expositions. However, at intermediate times, appearance of a new fluorescence band can be observed.

The Fig. 2a shows two luminescence spectra of the h-BN sample in ambient air before and after irradiation *under vacuum* (dose = 6 J/cm$^2$). In these low-dose conditions the surface darkening is not yet visible by a naked eye, however, the fluorescence spectrum is already modified. A new broad asymmetric band (labeled D) with a maximum at about 400 nm can be evidenced after the irradiation as shown by a difference spectrum.

The origins of this band can be now discussed. Fluorescence bands in the visible have been previously observed in thermally stimulated simultaneous luminescence and conduction measurements. They have been attributed to a free carriers recombination on deep impurity levels within the h-BN band gap related to carbon atom intercalated between two BN layers [28, 40]. In our case, the net result of the surface irradiation is the creation of multiple nitrogen vacancies until boron atoms appear. Keeping in mind the above assignment of the $V_N$ surface centers, we associate the deep levels responsible for the luminescent band D with nitrogen multiple vacancies m-$V_N$. According to ref [11], both 2-$V_N$ and 3-$V_N$ vacancies show electronic transitions in the UV spectral range of ~3 eV. The strongest $E_2(A \to M)$ transition of 2-$V_N$ is predicted slightly blue-shifted (~0.4 eV) with respect to that of 3-$V_N$. This assignment agrees with the progressive red shift of the visible emission band observed in our earlier experiments when the irradiation dose increases [38].

The oxygen gas produces another very interesting effect on irradiated h-BN samples. In contrast to samples irradiated under vacuum, no sample darkening due to elemental boron has been observed in samples irradiated under air. Even under high irradiation dose above 400 J/cm² ($F_L$ = 207 mJ/cm²) the sample surface conserved its original white color. Moreover, the luminescence spectrum is not changed after irradiation, which is shown in Fig. 2b. This is a pure effect of oxygen, while samples irradiated under a pure nitrogen atmosphere were strongly colored as this is observed under vacuum.

This effect may be explained by a formation of the stable oxide layer on the h-BN surface. As we have discussed above, surface $V_N$ vacancies are not capable to chemisorb molecular oxygen. UV laser photons break B-N bonds and multiple $V_N$ vacancies appear as irradiation progresses. In this process the surface acquire progressively a metallic nature and the attachment of oxygen becomes dissociative [41]. The boron in excess reacts with the oxygen atoms to form a protective layer of boron oxide on the surface:

$$xB(s) + \frac{y}{2}O_2(g) \rightarrow B_xO_y(s) \quad (7)$$

This boron oxide layer is transparent in the UV-visible spectral range [42] allowing laser photons to penetrate the sample bulk. However as we have discussed, only surface B-N bonds can be broken. Because all surface m-$V_N$ centers are healed by oxygen, no more boron production is now possible. The oxygen passivation prevents the B-N dissociation and preserves the sample stoichiometry.

This mechanism assumes a very fast formation of the oxide layer. Indeed, in ambient atmosphere the oxygen gas is readily physisorbed on $V_N$ centers. It is therefore available for reaction (7) when this center is transformed into m-$V_N$. Moreover even when the appeared 3-$V_N$ center was free of adsorbed oxygen molecule at the moment of transformation, a new adsorption event will take place before the next laser photon produces boron atom. Indeed, using the oxidation rate $k_{vac} = 1.5 \cdot 10^{-4}$ s$^{-1}$ obtained under vacuum of 10$^{-7}$ mbar, [38] one can estimate the time required for surface oxidation at the atmospheric conditions (1 bar, ~20% O₂) as $\tau = 0.6$ μs. This time is much shorter than the laser pulses periodicity t = 0.1 s.

*UV2 band*

Below we discuss the nature of the structured UV2 band. Typical room temperature photoluminescence spectra of h-BN excited at 248 nm (5.0 eV) in air are shown in Fig.3. This band (300-360 nm) is structured and composed of four distinct peaks labeled (a), (b), (c) and (d). We have used multiple Gauss fit procedure to obtain the energetic position and intensity of each peak. Because of not high signal-to-noise ratio, we then averaged these values over twenty spectra that result in values of 304.3 nm (a), 317.1 nm (b), 331.6 nm (c) and 351.1 nm (d). The spacing between the four observed peaks is found regular. We have attributed the peaks (b), (c), and (d) to the phonon replicas of the band (a), with respectively 1, 2, and 3 emitted phonons. The plot of the energy of each peak against the number of phonons is shown in the inset of Fig. 4. A linear fit, taking into account peaks (a) (b) and (c), results in the phonon energy of ~168 ± 2 meV (1353 ± 14 cm$^{-1}$). This value is in agreement with the in-plane transversal phonon TO$_{//}$ : $\omega_{TO}$=1367 cm$^{-1}$, which is Raman active [9, 10]. The peak (d) deviates slightly with respect to the previous fit. This could be due to both low spectral resolution and/or potential anharmonicity of the interaction between the lattice and an impurity centre. The intensity distribution between four sub-bands (a)-(d) suggests a strong phonon coupling.

The luminescence band UV2 is known in literature [6, 8, 28, 35, 36, 38] but its origin subjected to discussions. Its assignment to a direct band to band transition has been proposed in Refs [6, 36], due to a linear increase of the emission intensity with the cw laser power [36], or to the invariance of the spectra after heating of samples in different gas environments [6]. On the other hand, based on electron paramagnetic resonance, thermoluminescence, thermally-stimulated current measurements [28], and semi-empirical molecular orbital calculations [43], Katzir *et al*. have assigned the luminescence to transitions between conduction band and impurities involving carbon substitutional nitrogen defect (eA$^0$ luminescence).

If the luminescence band UV2 is assisted to structural defects or impurities, those limited concentration in the h-BN sample will result in fluorescence intensity saturation with an increase of the laser fluence. No effect of laser fluence has been previously reported under vacuum in nanosecond pulse mode [38]. However, in ambient air neither strong fluorescence from multiple nitrogen vacancies (dominant bands V and D) no degradation of h-BN samples appear. In these conditions relatively weak UV2 band becomes dominant and can be carefully observed. Moreover, in ambient air the h-BN fluorescence lineshape does not changes with laser dose. In the same time, laser fluence affects these spectra. As shown in Fig.3, the UV1 band intensity keeps almost constant, whereas intensity of the UV2 band increases with the fluence. Moreover, at fluences above 100 mJ/cm$^2$ the saturation is observed. This behavior is characteristic of the impurity or defect centers involved in the PL process.

In the hypothesis of luminescence due conduction band – impurity atom transition proposed in [28, 43], the saturation of the UV2a band can be readily describe by a simple kinetics model. Assuming a limiting concentration of impurity atoms ($N_C$), a concentration $n$ of electrons after excitation by UV-laser photons, and their decay through dominant non-radiative and minor radiative (eA° luminescence) channels, the kinetic equation can be written as:

$$\frac{dn}{dt} = \frac{\sigma F_L}{h\nu \tau_L}(N_C - n) - \frac{1}{\tau}n \qquad (8)$$

where $F_L$ is the laser fluence, $\sigma$ is the photoabsorption cross section of the impurity at $h\nu$ = 5 eV, $\tau_L$ is the pulse duration and $\tau$ is the electron lifetime. The excitation time $\tau_L$ = 15 ns is much longer than the fluorescence lifetime $\tau$ which is less than 1 ns for the UV2 band [31, 36]. This allows for steady-state solutions of rate equations (9)

$$I_{UV2a} \propto n = \frac{F_L}{F_L + \bar{\bar{F_L}}} \cdot N_C \qquad (9)$$

with $\overline{F_L} = \frac{h\nu}{\sigma} \cdot \frac{\tau_L}{\tau}$. According to (9), $F_L = \overline{F_L}$ corresponds to the condition $n = N_C/2$. The fit of the experimental data in Fig. 4 by Eq. (9) with $\overline{F_L} = 35$ mJ/cm² is shown by solid line. From this value and taking a lifetime $\tau$ = 1 ns, we can estimate the photoabsorption cross section of the impurity $\sigma = 3.4 \cdot 10^{-16} \, cm^2$ at 5 eV. This value will allow to obtain the impurity concentration from absorbance measurements at 248 nm.

Our results confirm that impurities are involved in the UV2 fluorescence. Their chemical nature is not yet clear. We can not exclude participation of carbon impurities in the fluorescent process. More experimental series with samples of known carbon doping may provide a conclusive picture of a role of carbon impurities in hBN luminescence.

## IV. Conclusion

Photoluminescence (PL) of hexagonal nitride boron (h-BN) has been studied with nanosecond UV laser irradiation (248 nm) below ablation threshold in different environments: under vacuum, nitrogen atmosphere, and in ambient air. Laser fluence and irradiation dose affect the PL spectra. The observed bands are tentatively ascribed to single and multiple nitrogen vacancies: $V_N$, 2-$V_N$, and 3-$V_N$. The band at 340 nm assigned to surface $V_N$ vacancies is observed under vacuum and quenched by physisorbed molecular oxygen. UV-irradiation of h-BN under vacuum results in a broad fluorescence band with maximum at 400 nm assigned to multiple $V_N$ vacancies; further irradiation breaks B-N bonds producing boron-enriched surface. However, no elementary boron appears under irradiation of samples in ambient atmosphere. This effect is explained by the oxygen healing of radiation-induced surface defects. Formation of oxide layer prevents the B-N dissociation and preserves the bulk sample stoichiometry. The structured fluorescence band between 300 nm and 350 nm is assigned to band – impurity transitions and its complex lineshape is explained by TO$_{//}$-phonon replica ($\omega$=168 ±2 meV).

# Acknowledgments

The experiments, presented in this paper, were performed at the Ultraviolet Laser Facility operating at IESL-FORTH, Heraklion, Crete, Greece and have been supported by the European Commission through the Research Infrastructures activity of FP6 (Project: "Laserlab-Europe", Contract: RII3-CT-2003-506350). The authors are particularly grateful to A. Egglezis for providing technical support during the experiments and to W. Marine and C. Fotakis for helpful discussions.

# References

1.  O.Mshima and K.Era, *Science and technology of boron nitride*. Electronic and Refractory materials, ed. Y.Kumashiro. 2000: Yokohama National University, Japan. 495-556.
2.  Huang, J.H. and Y.T. Zhu. *Advances in the synthesis and characterization of boron nitride*. in *Defect and Diffusion Forum*. 2000.
3.  Watanabe, K., T. Taniguchi, and H. Kanda, *Ultraviolet luminescence spectra of boron nitride single crystals grown under high pressure and high temperature.* Physica status solidi (a), 2004. **201**(11): p. 2561-2565.
4.  Watanabe, K., T. Taniguchi, and H. Kanda, *Direct bangap properties and evidence for ultraviolet lasing of hexagonal boron nitride single crystal.* Nature Materials, 2004. **3**: p. 404-409.
5.  Taylor, C.A., et al., *Observation of near-bandgap luminescence from boron nitride films.* Applied Physics Letters, 1994. **65**(10): p. 1251-1253.
6.  Larach, S. and R.E. Shrader, *Multiband luminescence in Boron Nitride.* Physical Review, 1956. **104**(1): p. 68.
7.  Larach, S. and R.E. Shrader, *Electroluminescence from boron nitride.* Physical Review, 1956. **102**(2): p. 582.
8.  Solozhenko, V.L., et al., *Bandgap energy of graphite-like hexagonal boron nitride.* Journal of Physics and Chemistry of Solids, 2001. **62**: p. 1331.
9.  Hoffman, D.M., G.L. Doll, and P.C. Eklund, *Optical properties of pyrolityc boron nitride in the energy range 0.05-10 eV.* Physical review B, 1984. **30**(10): p. 6051.
10. Geick, R., C.H. Perry, and G. Rupprecht, *Normal modes in hexagonal boron nitride.* Physical Review, 1966. **146**(2): p. 543.
11. Grinyaev, S.N., et al., *Optical absorption of hexagonal boron nitride involving nitrogen vacancies and their complexes.* Physics of the solid state, 2004. **46**(3): p. 435-441.
12. Chen, G., et al., *Optical absorption edge characteristics of cubic boron nitride thin films.* Applied Physics Letters, 1999. **75**(1): p. 10.
13. Mansour, A. and S.E. Schnatterly, *Anisotropy of BN and Be x-ray emission bands.* Physical Review B, 1987. **36**(17): p. 9234.
14. Carson, R.D. and S.E. Schnatterly, *X-ray emission from core excitons.* Physical Review letters, 1987. **59**(3): p. 319.
15. MacNaughton, J.B., et al., *Electronic structure of boron nitride single crystals and films.* Physical review B, 2005. **72**: p. 195113.
16. Carlisle, J.A., et al., *Band-structure and core-hole effects in resonnant inelastic soft X-ray scattering: Experiment and theory.* Physical Review B, 1999. **59**(11): p. 7433.

17. Jia, J.J., et al., *Resonant inelastic X-ray scattering in hexagonal boron nitride observed by soft X-ray fluorescence spectroscopy.* Physical Review letters, 1996. **76**(21): p. 4054.
18. Davies, B.M., et al., *Core excitons at the boron K edge in hexagonal BN.* Physical review B, 1981. **24**(6): p. 3537.
19. Brown, F.C., R.Z. Bachrach, and M. Skibowski, *Effect of X-ray polarization at the boron K edge in hexagonal BN.* Physical review B, 1976. **13**(6): p. 2633.
20. Tarrio, C. and S. Schnatterly, *Interband transitions, plasmons and dispersion in hexagonal boron nitride.* Physical review B, 1989. **40**(11): p. 7852.
21. Leapman, R.D., P.L. Fejes, and J. Silcox, *Orientation dependence of core edges from anisotropic materials determined by inelastic scattering of fast electrons.* Physical review B, 1983. **25**(5): p. 2361.
22. Leapman, R.D. and J. Silcox, *Orientation dependence of core edges in electron-energy-loss spectra from anisotropic materials.* Physical Review letters, 1979. **42**(20): p. 1361.
23. Blase, X., et al., *Quasiparticle band structure of bulk hexagonal boron nitride and related systems.* Physical review B, 1995. **51**: p. 6868.
24. Furthmüller, J., J. Hafner, and G. Kresse, Physical review B, 1994. **50**: p. 15606.
25. Xu, Y.N. and W.Y. Ching, Physical review B, 1991. **44**: p. 7787.
26. Liu, L., Y.P. Feng, and Z.X. Shen, *Structural and electronic properties of h-BN.* Physical Review B, 2003. **68**: p. 104102.
27. Arnaud, B., et al., *Huge excitonic effects in layered hexagonal boron nitride.* Physical Review letters, 2006. **96**: p. 026402.
28. Katzir, A., et al., *Point defects in hexagonal boron nitride. I. EPR, thermoluminescence and thermally-stimulated-current measurements.* Physical Review B, 1975. **11**(6): p. 2370.
29. Lukomskii, A.I., V.B. Shipilo, and L.M. Gameza, *Luminescence properties of graphite-like boron nitride.* Journal of applied spectroscopy, 1993. **57**(1-2): p. 607.
30. Kawaguchiaki, M., Y. Kita, and M. Doi, *Photoluminescence characteristics of BN(C,H) prepared by chemical vapour deposition.* Journal of materials science, 1991. **26**: p. 3926.
31. Era, K., F. Minami, and T. Kuzuba, *Fast luminescence from carbon-related defect of hexagonal boron nitride.* Journal of luminescence, 1981. **24/25**: p. 71.
32. Yao, B., et al., *Effects of degree of three dimensional order and Fe impurities on photoluminescence of boron nitride.* Journal of applied physics, 2004. **96**(4): p. 1947.
33. Yao, B., et al., *Strong deep-blue photoluminescence of mesographite boron nitride.* Journal of Physics: Condensed Matter, 2004. **16**: p. 2181.

34. Chen, Y., et al., *Boron nitride nanotubes:Pronounced resistance to oxidation.* Applied Physics Letters, 2004. **84**(13): p. 2430 - 2432.
35. Berzina, B., et al., *Photoluminescence excitation spectroscopy in boron nitride nanotubes compared to mycrocrystalline h-BN and c-BN.* Physica status solidi (c), 2005. **2**(1): p. 318.
36. Wu, J., et al., *Raman spectroscopy and time resolved photoluminescence of BN and $B_xC_yN_z$ nanotubes.* Nano Letters, 2004. **4**: p. 647.
37. Michel, J.P., et al., to be published, 2006.
38. Kanaev, A.V., et al., *Femtosecond and ultraviolet laser irradiation of graphite-like hexagonal boron nitride.* Journal of Applied Physics, 2004. **96**(8): p. 4483-4489.
39. Hirayama, Y. and M. Obara, J. Appl. Phys., 2001. **90**: p. 6447.
40. Lopatin, V.V. and F.V. Konusov, *Energetic states in the boron nitride band gap.* Journal of Physics and Chemistry of Solids, 1992. **53**(6): p. 847-854.
41. Bäuerle, D., *Laser Processing and Chemistry*. 3 ed. 2000, Berlin Heidelberg: Springer-Verlag.
42. Li, D. and W.Y. Ching, *Electronic structures and optical properties of low- and high-pressure phases of crystalline $B_2O_3$.* Physical review B, 1996. **54**: p. 13616.
43. Zunger, A. and A. Katzir, *Points defects in hexagonal boron nitride. II Theoretical studies.* Physical review B, 1975. **11**(6): p. 2378.

Figure captions

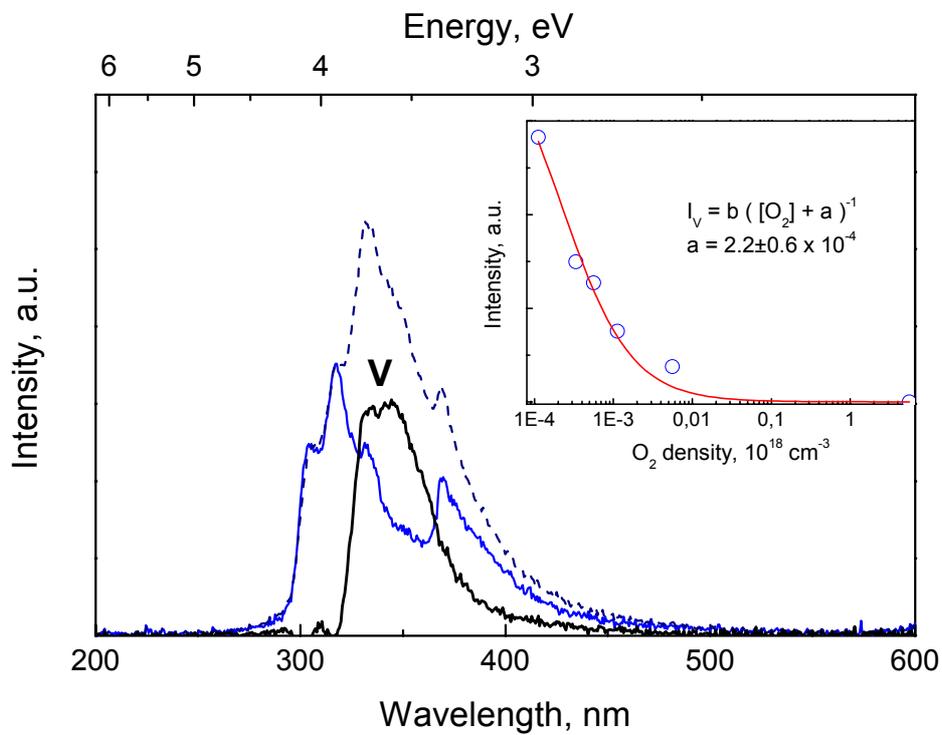

**Fig 1** Fluorescence spectra of h-BN in vacuum (dashed line) and in air (solid line). The laser fluence is $F_L$ = 3 mJ/cm$^2$ ($\lambda_{exc}$ = 248 nm). The band V is obtained by subtraction of two spectra. The insert shows the variation of the band V intensity as function of oxygen gas concentration.

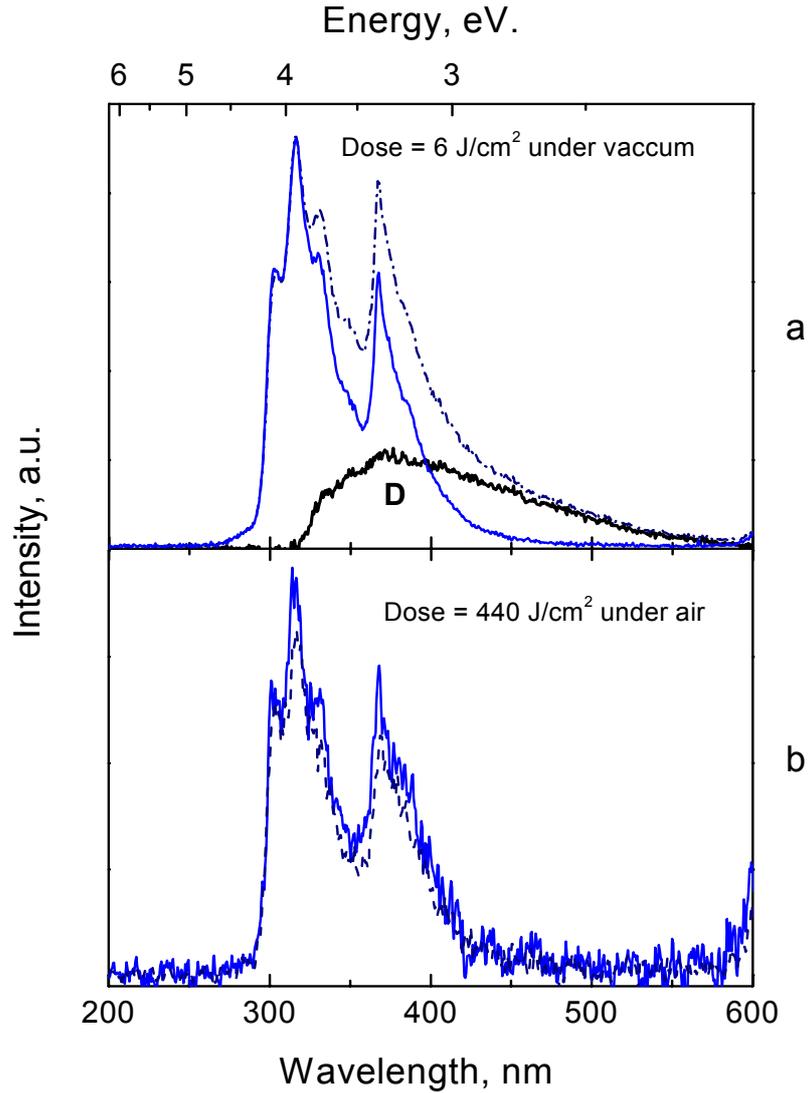

**Fig. 2** Luminescent spectra of h-BN in air obtained with laser fluence $F_L$ = 2.5 mJ/cm² ($\lambda_{exc}$ = 248 nm): (a) before (solid line) and after (dashed line) irradiation under vacuum with a dose of 6 J/cm²; (b) before (—) and after (- - -) irradiation in air with a dose of 440 J/cm². The band D is obtained by subtraction of two spectra. The difference in the signal to noise ratio between (a) and (b) is due to differences in the optical alignment between the quartz lens (that collect the luminescence) and the optical fiber (that guide the light to monochromator).

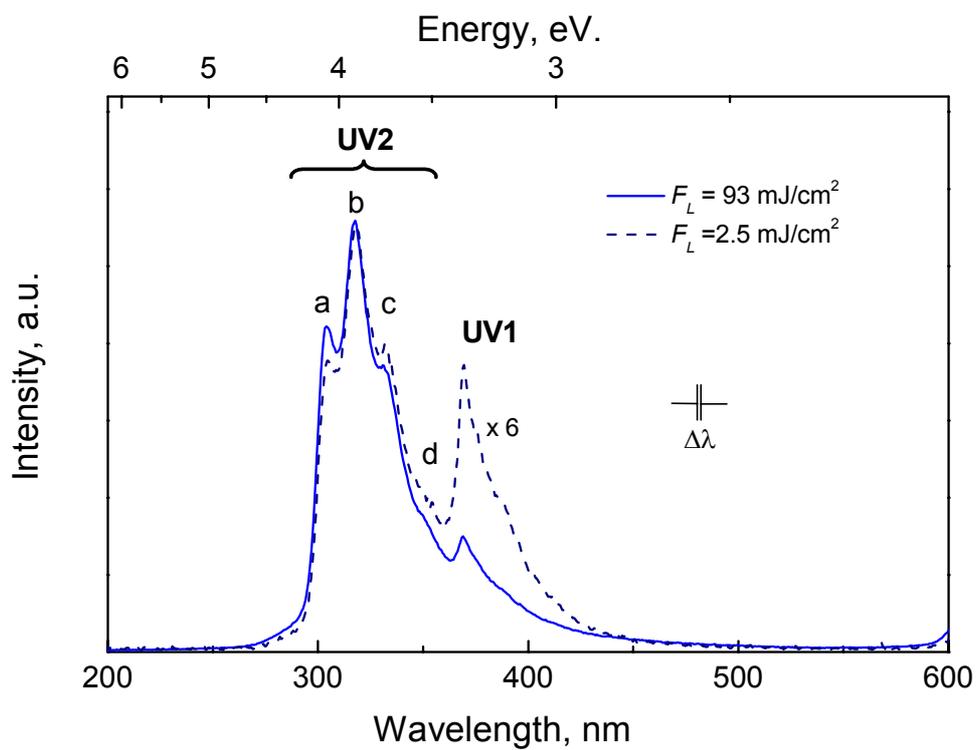

**Fig. 3** Fluorescence spectra of h-BN in air irradiated at laser fluence of 2.5 mJ/cm$^2$ (dashed line) and 93 mJ/cm$^2$ (solid line) ($\lambda_{exc}$ = 248 nm).

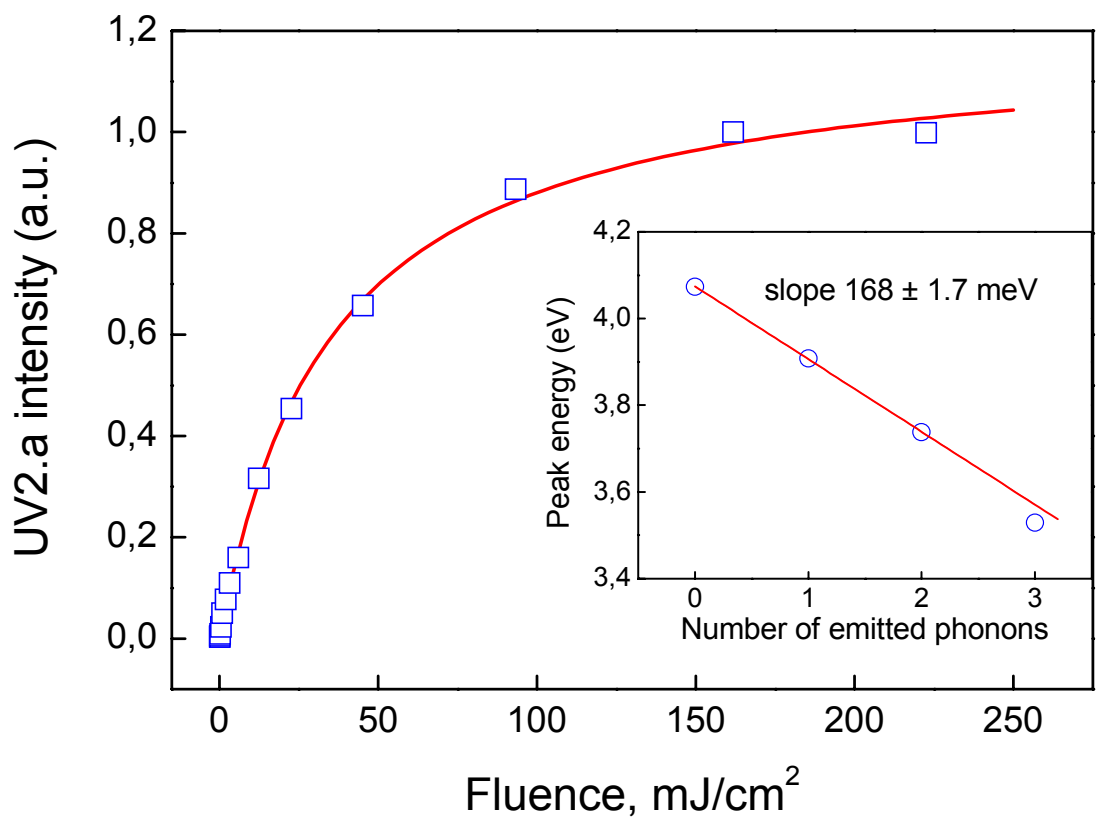

**Fig. 4** Evolution of the luminescent band UV2.a as function of the laser fluence ($\lambda_{exc}$ = 248 nm). In insert the energy of four peaks a-d of the UV2 band is plotted versus the number of emitted phonons. The slope obtained from a linear fit is also indicated.